\documentclass[preprint,12pt]{elsarticle}

\usepackage{amssymb}
\usepackage{graphicx}
\usepackage{dcolumn}
\usepackage{color}

\newcommand{\beq}{\begin{equation}}
\newcommand{\eeq}{\end{equation}}
\newcommand{\beqn}{\begin{eqnarray}}
\newcommand{\eeqn}{\end{eqnarray}}

\journal{Physics Letter B}
\begin{document}

\begin{frontmatter}

\title{The first candidate for chiral nuclei in the $A\sim80$ mass region: $^{80}$Br}

\author[SDU]{S.Y. Wang}
\author[SDU]{B. Qi}
\author[SDU]{L. Liu}
\author[PKU]{S.Q. Zhang\corref{corr1}}\ead{sqzhang@pku.edu.cn}
\author[PKU]{H. Hua}
\author[PKU]{X.Q. Li}
\author[PKU]{Y.Y. Chen}
\author[BUAA]{L.H. Zhu}
\author[PKU,BUAA,Stell]{J. Meng\corref{corr1}}\ead{mengj@pku.edu.cn}
\author[Stell]{S.M. Wyngaardt}
\author[Stell]{\\P. Papka}
\author[Stell,iThemba,Ilorin]{T.T. Ibrahim}
\author[iThemba]{R.A. Bark}
\author[iThemba]{P. Datta}
\author[iThemba]{E.A. Lawrie}
\author[iThemba]{J.J. Lawrie}
\author[iThemba]{S.N.T. Majola}
\author[iThemba]{P.L. Masiteng}
\author[iThemba]{S.M. Mullins}
\author[Debrecen]{\\J. G\'{a}l}
\author[Debrecen]{G. Kalinka}
\author[Debrecen]{J. Moln\'{a}r}
\author[Debrecen]{B.M. Nyak\'{o}}
\author[Debrecen]{J. Tim\'{a}r}
\author[Debrecen2]{K. Juh\'{a}sz}

\author[FZD]{R. Schwengner}
%
\cortext[corr1]{Corresponding author}
\address[SDU]{Shandong Provincial Key Laboratory of Optical Astronomy and Solar-Terrestrial Environment, School of Space Science and
Physics, Shandong University at Weihai, Weihai 264209, China}
\address[PKU]{State Key Lab Nucl. Phys. {\rm\&} Tech., School of Physics, Peking University, Beijing 100871, China}
\address[BUAA]{School of Physics and Nuclear Energy Engineering, Beihang University, Beijing 100191, China}
\address[Stell]{Department of Physics, University of Stellenbosch, Matieland 7602, South Africa}
\address[iThemba]{iThemba LABS, 7129 Somerset West, South Africa}
\address[Ilorin]{Department of Physics, University of Ilorin,  PMB 1515, Ilorin, Nigeria}
\address[Debrecen]{Institute of Nuclear Research of the Hungarian Academy of Sciences (ATOMKI), H-4001 Debrecen, P.O.Box: 51, Hungary}
\address[Debrecen2]{Department of Information Technology, University of Debrecen, Egyetem t\'{e}r 1, Debrecen, Hungary}
\address[FZD]{Institut f\"ur Strahlenphysik, Helmholtz-Zentrum Dresden-Rossendorf, D-01314 Dresden, Germany}

\date{\today}

\begin{abstract}
Excited states of $^{80}$Br have been investigated via the
$^{76}$Ge($^{11}$B, $\alpha$3n) and $^{76}$Ge($^{7}$Li, 3n)
reactions and a new $\Delta I$ = 1 band has been identified which
resides $\sim$ 400 keV above the yrast band. Based on the experimental results and
their comparison with the triaxial particle rotor
model calculated ones, a chiral character of the two bands within the
$\pi g_{9/2}\otimes \nu g_{9/2}$ configuration is proposed, which
provides the first evidence for chirality in the $A\sim80$ region.
\end{abstract}


\end{frontmatter}

\section{Introduction}

The spontaneously broken chiral symmetry is a well-known phenomenon in
chemistry, biology and particle physics. Recently, this phenomenon
in atomic nuclei has attracted significant attention and intensive
discussion.

The first prediction of chiral effects in atomic nuclei was made by
Frauendorf and Meng in 1997~\cite{FM97}. They pointed out that the
rotation of triaxial nuclei may give rise to pairs of identical
$\Delta I$=1 bands with the same parity, which are called chiral
doublet bands~\cite{FM97}. Presently, candidates for chiral doublet
bands have been observed experimentally in about 25 cases of odd-odd
nuclei, odd-$A$ and even-even nuclei (see
review~\cite{Meng08,Meng10} and references therein). Thus far most
studies on nuclear chirality have focused on the mass
$A\sim$130~\cite{Starosta01,Koike01,Bark01,Zhu03,Koike03,GR03,Wang06}
and 100~\cite{Vaman04,Joshi04,Timar06} regions, although the
explanations for some of them (e.g., $^{134}$Pr and $^{104}$Rh) are
still under debate~\cite{Vaman04,Joshi04,Tonev06,Petrache06,
Suzuki08}. However, there is no reason to consider the nuclei in
$A\sim$130 and 100 mass regions as unique in terms of the nuclear
chirality. Recently, a pair of negative-parity partner bands in
$^{198}$Tl have been suggested as candidate chiral
bands~\cite{Lawrie08}. Thus, it is interesting to search for more
candidates in other mass regions with the fingerprints for chirality
in Refs.~\cite{Meng10,Wang07CPL} to show that these chiral symmetry
properties are of a general nature and not related only to a
specific nuclear mass region. Following the previous investigations,
the experimental fingerprints for chirality established in
$A\sim$100 and 130 mass regions, i.e., the near degeneracy of
spectra, the similar $B(M1)$ and $B(E2)$ values, and the staggering
of $B(M1)/B(E2)$ ratios, should be used for new mass region.
Meanwhile the state of the art theoretical approach should be used
to examine the triaxial deformation and the particle-hole
configuration favorable for chirality.

Chiral rotation has also been predicted to occur in the $A\sim$80
mass region~\cite{Dimitrov00}, where chiral doublet bands may be
formed involving the $\pi g_{9/2} \otimes \nu g^{-1}_{9/2}$
configuration. For odd-odd Br isotopes with $A\sim$ 80, the proton
Fermi level ($Z=35$) lies at the bottom of the $\pi g_{9/2}$
subshell, and with an increase of the neutron number $N$ to the
magic number 50, the neutron Fermi level approaches the top of the
$\nu g_{9/2}$ subshell. Therefore, odd-odd Br isotopes with larger
$N$ are more suitable for sustaining chiral geometry than those with
smaller $N$, unless the value of $N$ is close to magic number
$N=50$, where rotational structures are not easily developed.

Based on the above consideration, the odd-odd nucleus $^{80}$Br may
be considered as a good candidate for a chiral nucleus. Hence, it is
interesting to populate high-spin states of the odd-odd nucleus
$^{80}$Br and to search for chiral doublet bands.

\section{Experiments}

High-spin states of $^{80}$Br were populated using the
fusion-evaporation reaction $^{76}$Ge($^{11}$B,$\alpha$3n) at a
beam energy of 54~MeV at the iThemba LABS in South Africa. The
target consisted of a 1.8~mg/cm$^2$ $^{76}$Ge metallic foil with
4.0~mg/cm$^2$ Au backing. The in-beam $\gamma-$rays were detected by
the AFRODITE array~\cite{JF04}, which consists of nine Compton-suppressed
clover detectors. The clover detectors have been arranged
in two rings at 90$^{\circ}$ and 135$^{\circ}$ with respect to the
beam direction, hence the directional correlation of oriented (DCO)
nuclei ratios for $\gamma$ transitions can be extracted. In
addition, the CsI particle detector array ---
Chessboard~\cite{Komati05} was also used with the AFRODITE array to
select specific reaction channels. A total of 2$\times$10$^8$
$\gamma - \gamma$ coincidence events were accumulated. In order to
double-check the experimental results, we have also used high-fold
coincidence $\gamma$-ray data from a measurement of the reaction
$^{7}$Li + $^{76}$Ge using an arrangement of six EUROBALL CLUSTER
detectors~\cite{JE92}. In that experiment, approximately
1.2$\times$10$^9$ coincidence events of fold 2 or higher were
collected and sorted off-line into $E_\gamma - E_\gamma$ matrices as
well as an $E_\gamma - E_\gamma - E_\gamma$ cube. The odd-odd
nucleus $^{80}$Br was produced as a by-product. More details on that
experiment can be found in Ref.~\cite{RS02}.

\begin{figure}[th]
\centering
\includegraphics[width=8cm, bb=200 120 550 500]{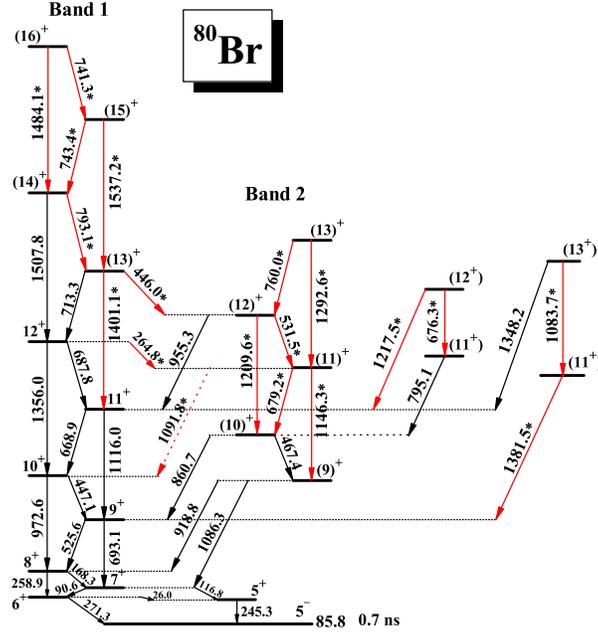}
\caption{\label{fig:scheme} (color online) Partial level scheme for
$^{80}$Br obtained in the present work. New observed transitions are
indicated by stars and red lines. }
\end{figure}

\section{Result and Discussions}

\begin{figure}[th]
 \centering
\includegraphics[angle=-90, scale=0.33, bb=160 40 455 740]{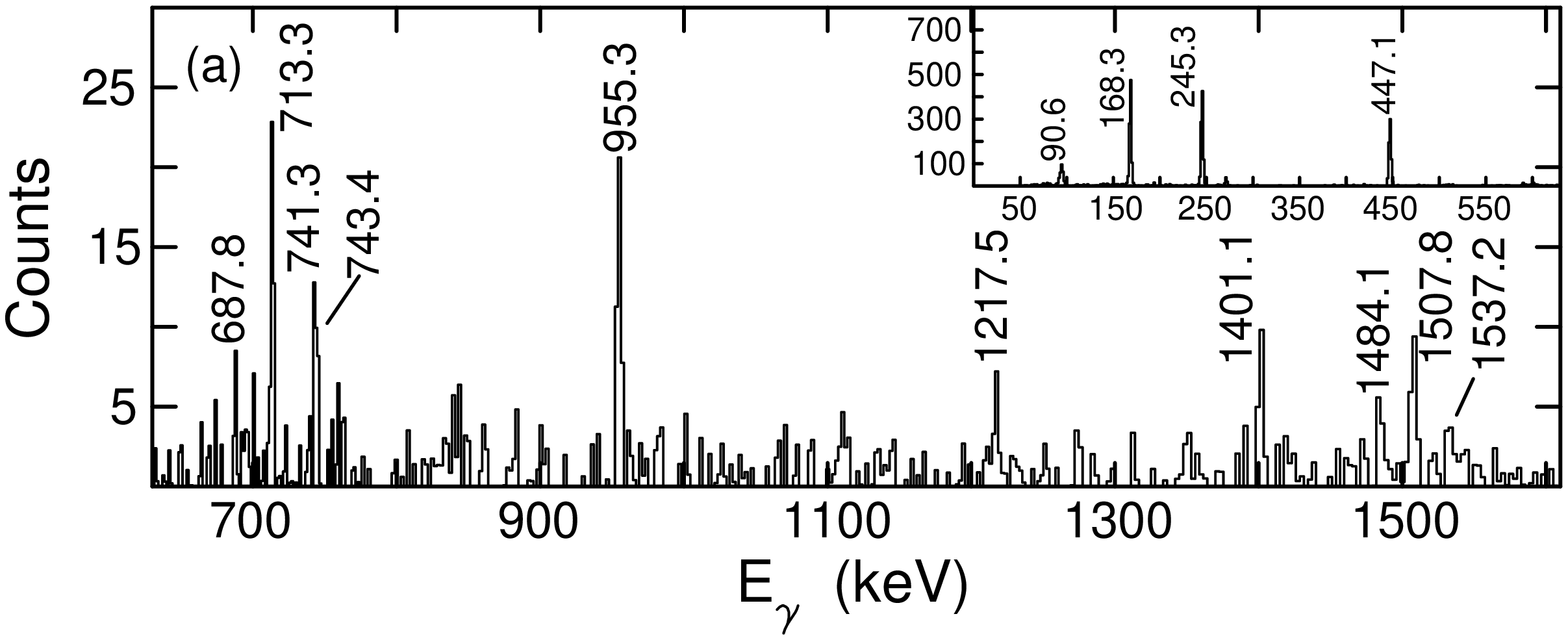}
\includegraphics[angle=-90, scale=0.33, bb=160 40 455 740]{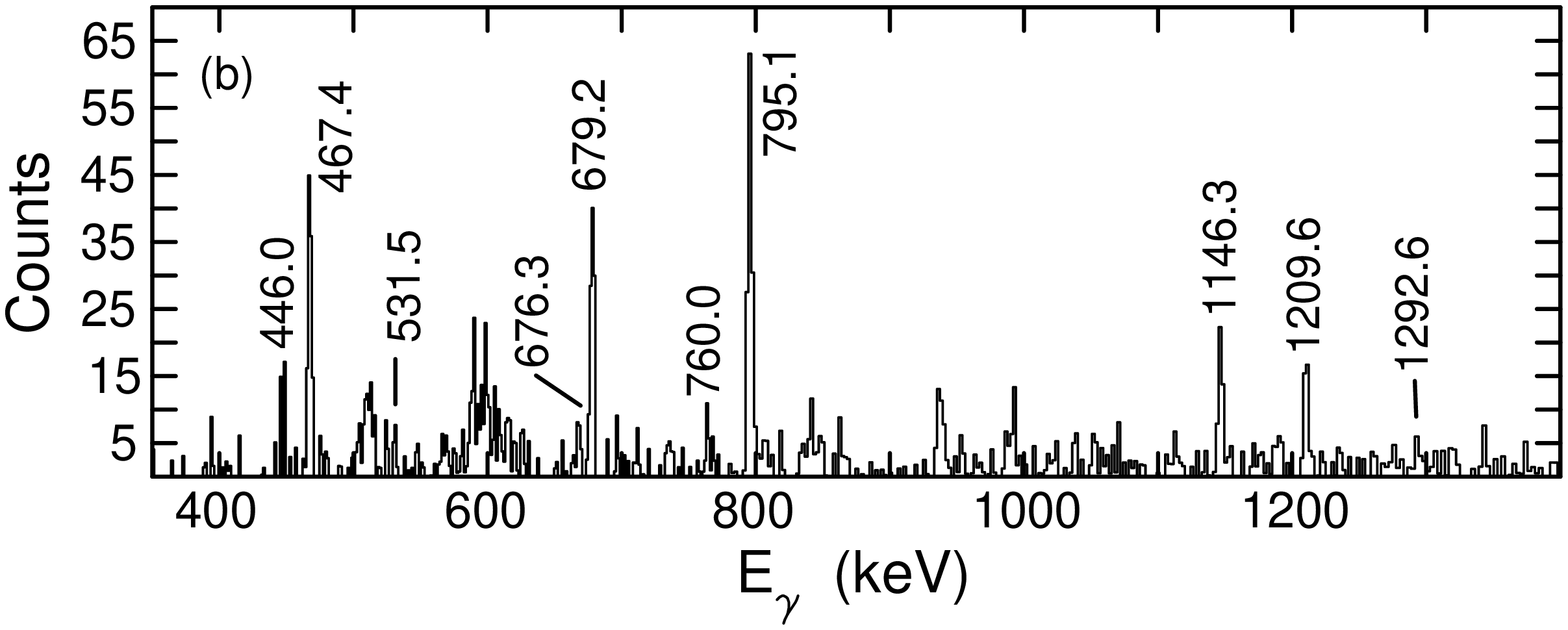}
\caption{Background subtracted coincidence spectra, revealing the
new transitions for band 1 (a) and band 2 (b) in $^{80}$Br. The
spectrum for band 1 is the sum of double gates: 1356.0\&525.6 and
668.9\&525.6 and the insert shows the low energy spectrum.
The spectrum for band 2 is the sum of double gates:
918.8\&168.3, 918.8\&467.4, 918.8\&1146.3, and 860.7\&525.6.}
\label{fig:spectra}
\end{figure}

A partial level scheme of $^{80}$Br derived from the present work
and typical doubly gated coincidence spectra extracted from the cube
are shown in Figs.~\ref{fig:scheme} and ~\ref{fig:spectra},
respectively. As shown in Fig.~\ref{fig:scheme}, the previously
reported positive-parity band structures~\cite{Ray00} have been
considerably extended, and about 20 new transitions were added to
the level scheme. The yrast band (labeled as Band 1) has been
extended up to $I^\pi$=(16)$^+$. Note that a 1401-keV $\gamma$-ray
instead of a 1348-keV one in Ref.~\cite{Ray00} was assigned to
transition $13^+ \rightarrow 11^+$ of the yrast band, which could be
verified by the coincident relationship of the 1401-keV $\gamma$-ray
seen from Fig.~\ref{fig:spectra}(a). In addition, a new $\Delta I
=1$ positive-parity band structure (labeled as Band 2) was
established up to spin (13$^+$). Remarkably, several linking
transitions between the two bands, not only from band 2 to band 1
(such as 860.7 and 955.3 keV) but also from band 1 to band 2 (264.8 and
446.0 keV), were observed in the present work.

In the level scheme construction, we adopted the spins and parities
for band 1 from Ref.~\cite{Ray00} and determined the spins and
parities of the other level sequences on the basis of the
multipolarities of transitions connecting them to this band. The
lowest energy state of band 2 decays by a 918.8-keV transition to
the 8$^+$ and by a 1086.3-keV transition to the 7$^+$ state of band
1 (see Fig.~1). These two linking transitions had been reported in
the previous work~\cite{Ray00}, where the 918.8-keV transition was
assigned to be a $\Delta I=1$ M1/E2 transition, and the 1086.3-keV
transition was proposed to have $\Delta I = 2$ E2 or $\Delta I = 0$
M1/E2 multipolarities. According to the general yrast argument that
levels populated in heavy-ion reactions usually have spins
increasing with increasing excitation energy~\cite{ZhengY04}, we
adopted the $\Delta I$ = 2 E2 character for the 1086.3-keV
transition. Thus the spin/parity assignment (9$^+$) has been made
for the lowest state of band 2. The multipolarity information on the
918.8- and 1086.3-keV transitions could be obtained accurately in
the second experiment. In that experiment, with setting gates on
stretched quadrupole transitions, the DCO ratios are $\sim$ 1 for
stretched quadrupole transitions and $\sim$ 0.5 for pure dipole
ones~\cite{RS02}. The present DCO ratios were extracted to be $\sim$
0.85 for the 1086.3-keV transition and $\sim$ 0.38 for the 918.8-keV
transition. This provides further support for the 9$^+$ assignment
for the lowest state of band 2. Furthermore, when the $\gamma$
transitions form rotational band structures, the cross-over
transitions were set to have the E2 character, while the normal
intra-band transitions were assumed to have M1/E2 nature. These
arguments lead to the current spin-parity assignments shown in
Fig.~1.

\begin{figure}[ht]
\centering
\includegraphics[width=6cm, bb=200 20 520 540]{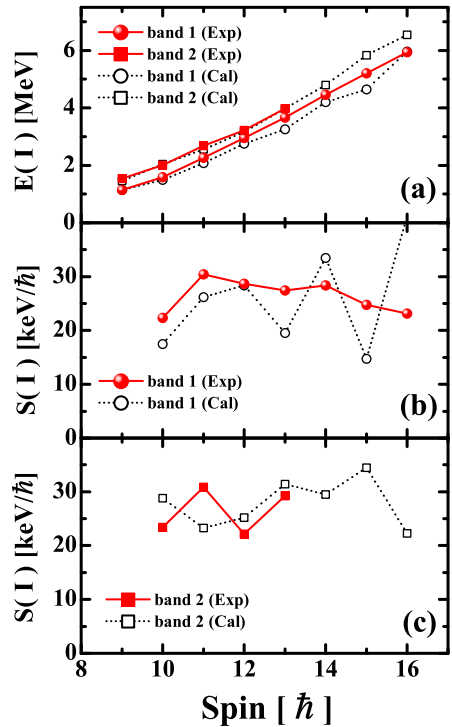}
\caption{(color online) (a) Excitation energy and (b,
c) staggering parameter S(I) = $[E(I)-E(I-1)]/2I$ as a function of spin
for the doublet bands in $^{80}$Br. The filled (open) symbols
connected by solid (dotted) curves denote experimental (theoretical)
values. The bands 1 and 2 are shown by circles and squares,
respectively.}\label{fig:EI}
\end{figure}

\begin{figure}[ht]
\centering
\includegraphics[width=6cm, bb=200 30 530 390]{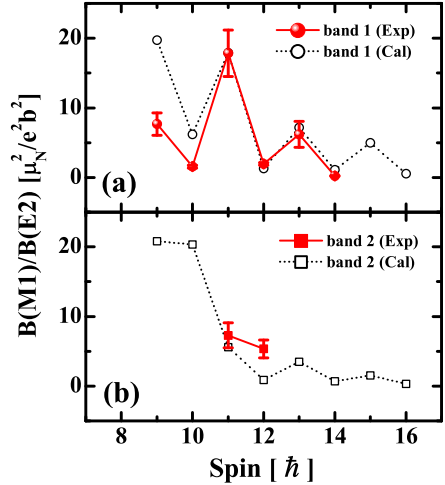}
\caption{(color online) Comparisons of the measured and calculated in-band $B(M1)/B(E2)$ ratios for the bands 1 (a) and 2 (b) in $^{80}$Br.}\label{fig:BME}
\end{figure}

Band 1 has been already assigned to the $\pi g_{9/2}\otimes \nu
g_{9/2}$ configuration~\cite{Ray00}. As shown in
Fig.~\ref{fig:scheme}, band 2 has the same parity as the band 1.
Furthermore, the existence of several M1/E2 and E2 linking
transitions between the two bands indicates that band 2 has the same
configuration $\pi g_{9/2}\otimes \nu g_{9/2}$ as that of the yrast
band, as discussed in Refs.~\cite{Starosta01,Koike01,GR03}. The
degree of degeneracy of bands 1 and 2 in $^{80}$Br is exhibited by
the excitation energies as a function of spin as shown in
Fig.~\ref{fig:EI}(a). The two bands maintain an energy difference
around $\sim$400 keV within the observed spin interval. The
experimental $B$(M1)/$B$(E2) ratios are also extracted and presented
in Fig.~\ref{fig:BME} for bands 1 and 2 in $^{80}$Br. As shown in
Fig.~\ref{fig:BME}, the $B$(M1)/$B$(E2) ratios for the two bands are
comparable in magnitude, and in particular, the ratios for band 1
show clearly the odd-even staggering as a function of spin. Taking
the experimental observations given above into account, bands 1 and
2 in $^{80}$Br may be considered as candidates for chiral doublet
bands, which had already been systematically reported in the odd-odd
nuclei of $A\sim$100 mass region with $\pi g_{9/2}\otimes \nu
h_{11/2}$ and of $A\sim$130 with $\pi h_{11/2}\otimes \nu h_{11/2}$.
An interpretation of $\gamma$-vibration coupled to the yrast band is
unlikely as the $\gamma$-vibration energies are larger than 600 keV
in this mass region~\cite{RS87, Wells80}.

In the following, the calculations of the triaxial relativistic
mean-field (RMF) theory~\cite{JM06} and the particle rotor model
(PRM)~\cite{Wang07,Zhang07,Wang08} are used to discuss whether band
2 could be interpreted as chiral partner of band 1, and to investigate to
the evolution of chiral geometry in this nucleus.

From the configuration-fixed constrained triaxial RMF
calculations~\cite{JM06} with parameter set PK1~\cite{Long04}, the
self-consistent deformation parameters $\beta_2$=0.346 and
$\gamma=24.59^\circ$ are obtained corresponding to the $\pi g_{9/2}
\otimes \nu g_{9/2}$ configuration in $^{80}$Br. These deformation
parameters are suitable for the construction of chiral bands.
Compared with the deformation values $\beta_2 \approx0.2 - 0.3$
needed to reproduce the chiral doublet bands in $A\sim 100, 130$
mass region, the value $\beta_2$=0.346 is obviously much larger. It
is noted that large quadrupole deformations of $\beta_2\approx0.3 -
0.4$ have been reported for the bands with $\pi g_{9/2} \otimes \nu
g_{9/2}$ configuration in the lighter $^{74,76}$Br
nuclei~\cite{Buccino90,Loritz99}. Lifetime measurements were carried
out for the $10^+$, $11^+$ and $12^+$ states in the yrast band of
$^{80}$Br in Ref.~\cite{Ray00}, and moderate quadrupole deformation
parameters $\beta_2=0.17,~0.12$, and 0.12 were deduced respectively
for these states with the assumption of the axial symmetry. If we
turn to adopt the essential triaxiality $\gamma=24.59^\circ$ from
the RMF calculation, according to the procedure in
Refs.~\cite{Petkov98,Andgren05}, a large quadrupole deformation
parameter $\beta_2=0.321$ will be determined for the yrast $10^+$
state of $^{80}$Br from its lifetime datum. In the triaxial deformed
case~\cite{Petkov98}, the reduced E2 transitional probability is
given by,
 \beq
 B(E2,I \rightarrow I-2)
 = \frac{5}{8
 \pi}Q_{0}^{2}\frac{(I-1)I}{(2I-1)(2I+1)}
 \left[\cos(\gamma+30^{\circ})-\cos(\gamma-30^{\circ})\frac{K^{2}}{(I-1)I}\right]^{2},
 \eeq
where the Lund convention for $\gamma$ is used and the intrinsic
quadrupole moment
 \beq
 Q_{0} = \frac{3}{\sqrt{5\pi}}Z R_{0}^{2}\beta_{2}(1+0.16\beta_{2})
 \eeq
with the nuclear radius $R_{0}=1.2 A^{1/3}$, whereas the projection
$K \sim 4$ is adopted here for the yrast band with the $\pi g_{9/2}
\otimes \nu g^{-1}_{9/2}$ configuration in $^{80}$Br. By this means,
the deformation parameters obtained from the triaxial RMF
calculations are consistent with the previous performed lifetime
measurement in this nucleus~\cite{Ray00}. We then adopted these
self-consistent RMF deformation parameters as the inputs to the
triaxial PRM calculations. The valence nucleon configuration space
is expanded in the $1g_{9/2}$ proton subshell and the $1g_{9/2}$
neutron subshell, and the proton and neutron Fermi surfaces for
$^{80}$Br are placed in the $\pi g_{9/2} 1/2$ and $\nu g_{9/2} 7/2$
orbitals, respectively. The pairing gaps $\Delta_p = 1.295$ MeV and
$\Delta_n = 1.287$ MeV are used for protons and
neutrons~\cite{Moller95}. As demonstrated in Ref.~\cite{Lawrie10},
restricting the configuration space to one orbital for the proton
and one for the neutron makes the calculated doublet bands to be
closer to each other, while larger configuration space is usually
more realistic.  For the electromagnetic transitions, the
gyromagnetic ratios $g_R = Z/A$ = 0.44, $g_p$ = 1.26, and $g_n =-
0.26$ are adopted. The moment of inertia $\Im = 12~\hbar^{2}$/MeV is
adjusted to the experimental energy spectra.

A comparison of the measured and calculated excitation energies
$E(I)$ and the energy staggering parameter $S(I)$, defined by
$[E(I)-E(I-1)]/2I$, is presented in Fig.~\ref{fig:EI}. One can see
that the PRM calculations reproduce the main features of the data
well. From Figs.~\ref{fig:EI}(b) and (c), the experimental values of
$S(I)$ yield an almost constant value of $\sim$25 keV$/\hbar$, which
consists with the expectation for ideal chiral doublet
bands~\cite{Vaman04}. At $I \geq 13 \hbar$, the theoretical $S(I)$
values overestimate the amplitude of staggering, which is due to the
large Coriolis effect at high spins. The attenuation of Coriolis
effect was not taken into account in the present PRM calculations.
In Fig.~\ref{fig:BME}, the calculated $B(M1)/B(E2)$ ratios for the
doublet bands are compared with experiment. It can be seen that the
agreement for the $B(M1)/B(E2)$ ratios at the whole spin region is
excellent. This staggering phase of doublet bands in $^{80}$Br is
similar to the case in the $A\sim$130 mass region with the $\pi
h_{11/2}\otimes \nu h_{11/2}$ configuration, while it is opposite to
the case for the $A\sim$100 mass region with the $\pi
g_{11/2}\otimes \nu h_{11/2}$ configuration.

\begin{figure}[ht]
\centering
\includegraphics[width=8cm, bb=50 50 780 720]{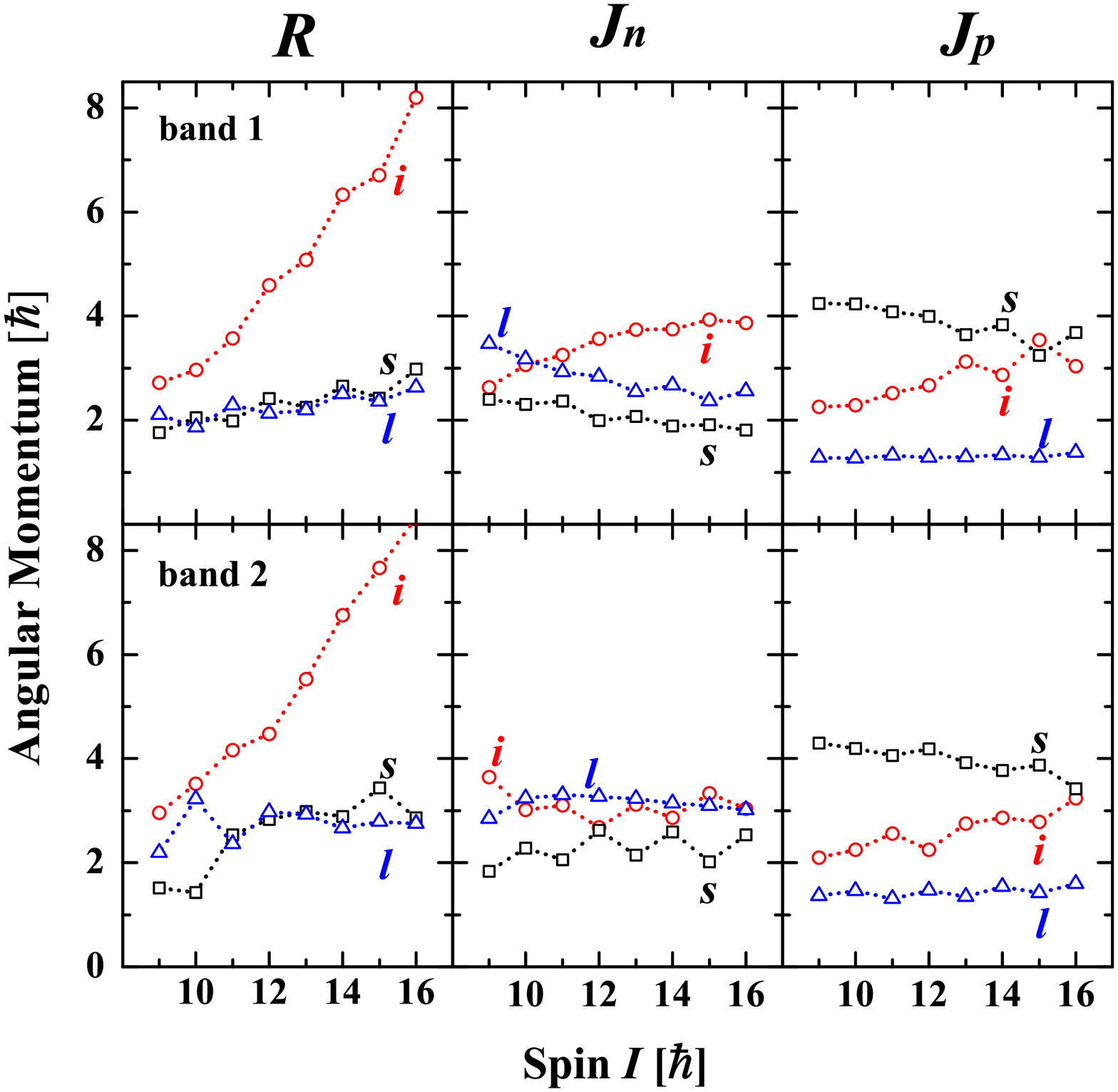}
\caption{(color online) The root mean square components along the
intermediate ($i$-, circles), short ($s$-, squares) and long ($l$-,
triangles) axis of the core, valence neutron, and valence protons
angular momenta calculated as functions of spin I by means of the
PRM for the doublet bands in $^{80}$Br.}\label{fig:AM}
\end{figure}

To exhibit the chiral geometry in $^{80}$Br, the expectation values
of the squared angular momentum components of the core ${R_i} =
\sqrt{\langle \hat{R}_{i}^2 \rangle}$, the valence neutron $J_{ni} =
\sqrt{\langle \hat{j}_{ni}^2\rangle}$, and the valence proton
$J_{pi}=\sqrt{\langle \hat{j}_{pi}^2\rangle}$, ($i=l, s, i$) for the
doublet bands are presented in Fig.~\ref{fig:AM}. The collective and
valence-proton angular momenta align along the intermediate axis,
and the short axis, respectively. For an ideal chiral geometry, the
angular momentum of the valence neutron is expected to align along the
long axis. As shown in Fig.~\ref{fig:AM}, however, the angular
momentum of the hole-like $g_{9/2}$ valence neutron ($g_{9/2}[413]
\frac{7}{2}$) in $^{80}$Br shows a large mixture between the long
and intermediate axes. Therefore, the present coupling pattern of
angular momenta somewhat departs from the ideal chiral geometry.
However, the total angular momentum is still aplanar.

\begin{figure*}[ht]
\centering
\includegraphics[width=14cm, bb=30 260 750 850]{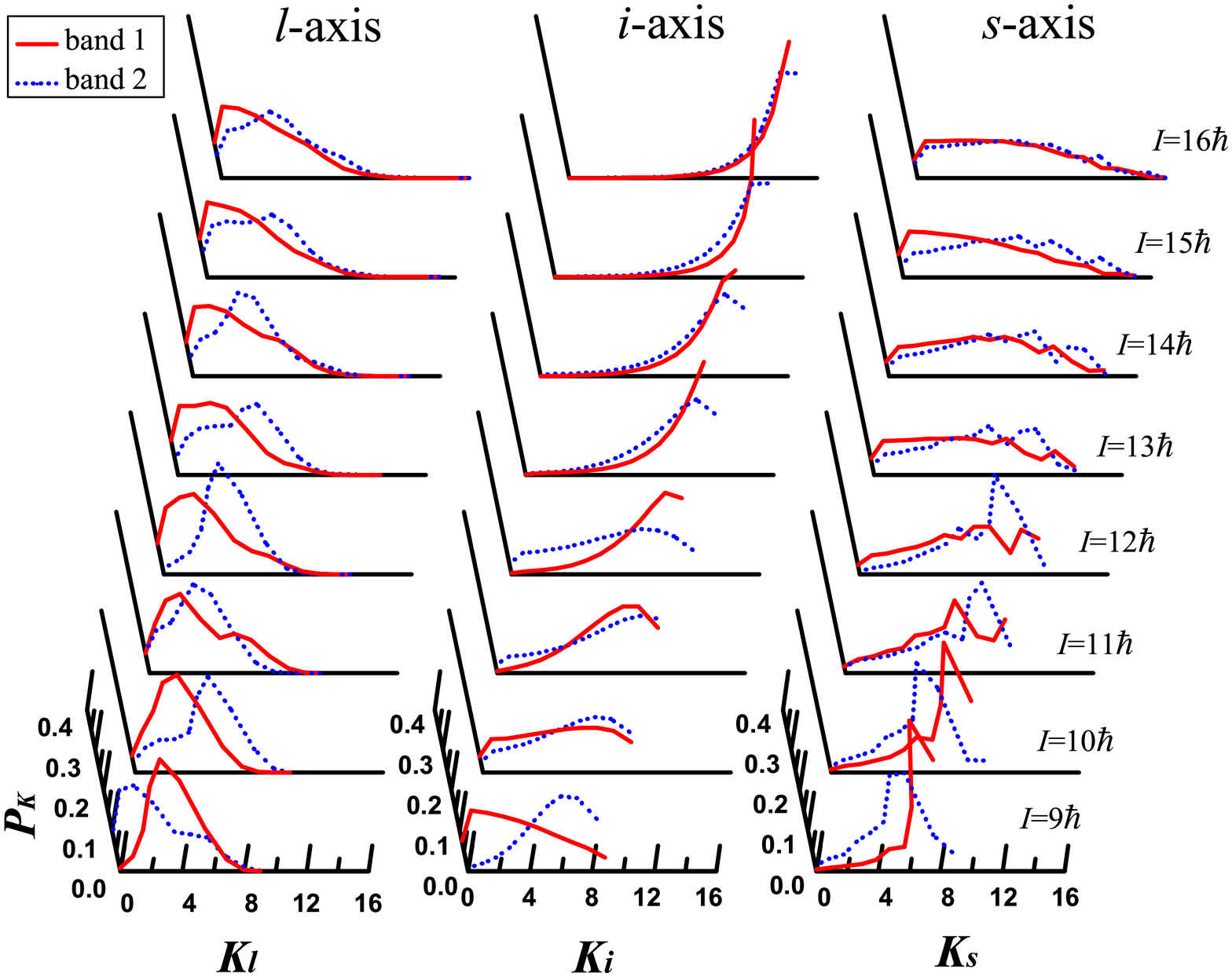}
\caption{\label{fig:K}. The probability distributions for projection
of total angular momentum on the long ($l$-), intermediate ($i$-)
and short ($s$-) axis in PRM for the doublet bands in $^{80}$Br.}
\end{figure*}

To give further understanding of the evolution of the chirality with
angular momentum~\cite{Qi0901}, the probability distributions for
the projection of the total angular momentum along the $l$-, $i$-
and $s$-axes are given in Fig.~\ref{fig:K} for the doublet bands in
$^{80}$Br. At the bandhead ($I=9~\hbar$), the probability
distribution of two bands differs as expected for a chiral
vibration~\cite{Qi0901,Mukhopadhyay07,Wang10}. For the yrast band
the maximum probability for the $i$-axis appears at $K_i\approx0$,
whereas the probability for the higher band 2 is zero at
$K_i\approx0$, having its peak at $K_i=6$, which indicates an
oscillation of the collective angular momentum vector $\bf{R}$
through the $s$-$l$-plane. The characteristics of chiral vibration
can be found also at other spins, and the pure static chirality
which has the identical K distributions for the doublet
bands~\cite{Qi0902} is absent in $^{80}$Br. Note that the motion of
chiral vibration is mixed by the static chirality for spin
$I=10~\hbar$, reflected by the fact that the $K_i$ distribution
probability in band 1 has a bump for lower $K_i$, while it is
similar as band 2 for higher $K_i$~\cite{Qi0902}. For higher spins,
another type of chiral vibration which contains oscillations along
both $l$ and $s$ axes can be seen. The lack of static chirality in
$^{80}$Br should be attributed to the Fermi surface of the neutron
placed in the $\nu g_{9/2}[413] \frac{7}{2}$ orbital instead of the
top of the $\nu g_{9/2}$ subshell, where less orthogonal coupling is
obtained for the angular momenta between the rotor $\bf{R}$ and the
neutron $\bf{J_n}$.

Along the states of band 2 up to spin 13$^+$, there are several
states with the same spin-parity: 11$^+$ at 2522 keV, 11$^+$ at 2797
keV, 12$^+$ at 3473 keV, and 13$^+$ at 3606 keV (see
Fig.~\ref{fig:scheme}). These states do not form an obvious
collective structure but they are connected with M1/E2 and E2
transitions either to the band 1 or band 2. Moreover, the energies
of these states are close to the corresponding states in band 2.
These experimental observations suggest that these states may be
built on the same configuration with band 2, reflecting the fact
that the chiral geometry of $^{80}$Br is not very stable. A similar
pattern of unstable chiral structure was observed and discussed in
$^{132}$Cs~\cite{GR03}. These observations are consistent with the
$K$ distribution patterns which are shown in Fig.~\ref{fig:K}, and
provide additional support for the interpretation of chiral
vibration for the doublet bands in $^{80}$Br. The unstable of chiral
structure in $^{132}$Cs is due to the moderate
deformation~\cite{GR03}, while the reason in $^{80}$Br may come from
the fact that the neutron Fermi surface is closer to  $\nu g_{9/2} 7/2$
orbital than to $\nu g_{9/2} 9/2$ orbital.

\section{Summary}

In summary, excited states of $^{80}$Br were investigated by means
of in-beam $\gamma$-ray spectroscopy. The previously known yrast
band has been extended up to $I^\pi$=(16)$^+$. A new $\Delta I$ = 1
band has been identified which resides $\sim$ 400 keV above the
yrast band. The experimental $B(M1)/B(E2)$ ratios extracted for the
two bands are comparable in magnitude, and in particular, the ratios
for band 1 show clearly the odd-even staggering. By examining the
experimental observations against the fingerprints for chirality in
Refs.~\cite{Meng10,Wang07CPL}, the two bands can be considered as
candidates for chiral doublet bands in the $A\sim$80 mass region.

The configuration-fixed constrained triaxial RMF approach is applied
to determine quadrupole deformations for the $\pi g_{9/2}\otimes \nu
g_{9/2}$ configuration in $^{80}$Br. Self-consistent deformation
parameters $\beta _2$ = 0.346 and $\gamma=24.59^\circ$ are obtained.
These are favorable deformation parameters for chirality. Using
these self-consistent deformation parameters from the RMF approach
as input, the present triaxial PRM calculation provides a good
description of the positive parity doublet bands in $^{80}$Br. The
analysis of the probability distributions of the angular momentum
indicates that the doublet bands in $^{80}$Br might correspond to a
typical chiral vibration pattern.

From studies of the odd-odd $^{80}$Br nucleus involving the $\pi
g_{9/2}\otimes \nu g_{9/2}$ configuration, we have found a new
region of chirality. The nucleus $^{80}$Br presents mainly the
characteristics of chiral vibration. Further efforts are needed to
explore nuclear chirality and search for the possible static pattern
in this $A\sim80$ region.

\section*{Acknowledgements}
This work is supported by the National Natural Science Foundation
(Grant Nos. 10775005, 10875002, 10875074, 10947013, 10975007,
10975008, 11005069 and 10710101087), the SA/CHINA research
collaboration in science and technology (Grant No. CS05-L06), the
Shandong Natural Science Foundation (Grant No.~ZR2010AQ005), and the
Major State Research Development Programme (No. 2007CB815005) of
China, the collaboration between the National Research
Foundation of South Africa (NRF, Contract Number UID61851) and the Hungarian
National Office for Research and Technology (NKTH, Contract Number
ZA-2/2008). One of the authors (K. Juh\'{a}sz) acknowledges support from the T\'{A}MOP 4.2.1./B-09/1/KONV-2010-0007/
IK/IT project, implemented through the New Hungary Development Plan.

We wish to thank the iThemba LABS technical staff and accelerator
group for their support and providing the beam.


\end{document}